\documentclass[conference]{IEEEtran}
\usepackage{booktabs} 

\usepackage{xspace}
\usepackage{multirow}
\usepackage{url,balance}
\usepackage{amsmath}
\usepackage{graphicx}
\usepackage{subfig}
\usepackage{epstopdf}
\usepackage{todo}
\usepackage{xifthen}
\usepackage{bbding}
\usepackage{dcolumn}
\usepackage[sort]{cite}
\usepackage{paralist}
\usepackage{color}
\usepackage[capitalize]{cleveref}
\usepackage{algorithm}
\usepackage[noend]{algorithmic}
\usepackage{tikz}
\usetikzlibrary{shapes, arrows, shadows, positioning}
\usepackage{amsthm}

\usepackage{color, colortbl}
\definecolor{Gray}{gray}{0.96}

\newcounter{mn}
\newcommand{\superscript}[1]{\ensuremath{{}^{\textrm{\scriptsize #1}}}}

\newcommand{\mntext}[1]{\colorbox{pink}{\begin{color}{black}#1\end{color}}}
\newcommand{\mn}[2][]{{\tiny\superscript{\mntext{\arabic{mn}}}}\marginpar{\scriptsize{
  \ifthenelse{\isempty{#1}}
  {\mntext{\parbox{0.95\marginparwidth}{\superscript{\arabic{mn}} \raggedright{#2}}}}
  {\mntext{\parbox{0.95\marginparwidth}{\superscript{\arabic{mn}}#1 says: \raggedright{#2}}}}
}}\stepcounter{mn}}

\newcommand{\cnsysname}{SilentWhispers\xspace}

\newcommand{\oursys}{SpeedyMurmurs\xspace}
\newcommand{\user}{user\xspace}
\newcommand{\users}{users\xspace}

\newcommand{\sender}{sender\xspace}
\newcommand{\receiver}{receiver\xspace}

\newcommand{\weight}{w}

\newcommand{\pathvar}{\var{p}}
\newcommand{\setC}{\textit{setCred}}

\newcommand{\setRoute}{\textit{setRoutes}}
\newcommand{\routePay}{\textit{routePay}}
\newcommand{\TA}{\mathcal{R}}

\newcommand{\cnode}{cnode}

\newcommand{\paysys}{PBT network\xspace}
\newcommand{\paysyss}{PBT networks\xspace}
\newcommand{\money}{funds\xspace}
\newcommand{\pathlength}{k}

\newcommand{\lmid}{l}
\newcommand{\lm}{L}

\newcommand{\id}{id}

\newcommand{\numT}{|\lm|}
\newcommand{\numA}{a}
\newcommand{\tl}{tl}
\newcommand{\epoch}{\var{epoch}}

\newcommand{\var}[1]{\textit{#1}\xspace}

\newcommand{\ith}{\ensuremath{i\superscript{\textrm \scriptsize th}}}

\newcolumntype{d}[1]{D{.}{.}{#1}}

\title{Settling Payments Fast and Private: Efficient Decentralized Routing for Path-Based Transactions}

\author{
\IEEEauthorblockN{Stefanie Roos}
\IEEEauthorblockA{University of Waterloo\\
sroos@uwaterloo.ca}
\and
\IEEEauthorblockN{Pedro Moreno-Sanchez}
\IEEEauthorblockA{Purdue University\\
pmorenos@purdue.edu}
\and
\IEEEauthorblockN{Aniket Kate}
\IEEEauthorblockA{Purdue University\\
aniket@purdue.edu}
\and 
\IEEEauthorblockN{Ian Goldberg}
\IEEEauthorblockA{University of Waterloo\\
iang@cs.uwaterloo.ca}
}

\begin{document}

\maketitle

\vspace{-1em}

\begin{abstract}
\footnote{This paper will appear at NDSS 2018} 
Decentralized path-based transaction (PBT)
networks maintain local payment channels between participants. Pairs of users 
leverage these channels to settle payments via a path of intermediaries
without the need to record all transactions in a global blockchain. 
PBT networks such as Bitcoin's Lightning Network and Ethereum's Raiden Network are the most prominent examples of
this emergent area of research. Both networks overcome scalability issues of widely used  
cryptocurrencies by replacing expensive and slow on-chain blockchain operations with inexpensive and fast off-chain transfers.

At the core of a decentralized PBT network is a routing algorithm that discovers transaction paths between sender and receiver.
 In recent years, a number of routing algorithms have been proposed, 
including landmark routing, utilized in the decentralized IOU credit network \cnsysname,  and Flare, a link state algorithm for the Lightning Network.
However, the existing efforts lack either efficiency or privacy, 
as well as the comprehensive analysis that is indispensable to ensure the success of PBT networks in practice. 
In this work, we first identify several efficiency concerns in existing 
routing algorithms for decentralized PBT networks.
Armed with this knowledge,
we design and evaluate \oursys , a novel routing algorithm for decentralized PBT networks 
using efficient and flexible embedding-based path discovery and on-demand 
efficient stabilization to handle the dynamics of a PBT network. Our simulation study,
based on real-world data from the currently deployed Ripple credit network, 
indicates that \oursys reduces the overhead of stabilization by 
up to two orders of magnitude and the overhead of routing a transaction by more than a factor of two. 
Furthermore, using \oursys maintains at least the same success 
ratio as decentralized landmark routing, while providing lower delays.
Finally, \oursys achieves key privacy goals for routing in decentralized PBT networks.   
\end{abstract}

\vspace{-1em}

\section{Introduction}
\label{sec:intro}

Since the advent of Bitcoin~\cite{nakamoto2011bitcoin}, many other blockchain-based 
payment systems have been proposed and deployed in practice to serve a multitude 
of purposes. For instance, IOweYou (IOU) credit networks~\cite{defigueiredo05trustdavis,Ghosh07} 
such as Ripple~\cite{ripple,armknecht2015ripple} 
or Stellar~\cite{stellar} leverage blockchain technology to enable 
real-time gross settlements~\cite{RTGS} between two end users  
across different currencies and assets significantly cheaper than the current central banking system. 
Ethereum~\cite{ethereum} builds on top of a blockchain 
to construct a platform to run fully expressive smart contracts.


However, the growing base of users and transactions is resulting in
blockchain scalability issues~\cite{Croman2016, poon2015bitcoin}. Moreover, the public nature of the blockchain 
leads to demonstrable privacy breaches of sensitive data such as the identities of the transaction partners and the transaction value~\cite{moreno2016listening, Meiklejohn2013, Kumar2017, Miller17,reid2013analysis}. 
Academic and industry efforts are leading towards peer-to-peer (P2P) %
path-based transaction (PBT) networks such as the Lightning Network~\cite{poon2015bitcoin} for Bitcoin, 
the Raiden Network~\cite{raiden} for Ethereum, 
SilentWhispers~\cite{malavolta17silent} for credit networks, or InterLedger~\cite{thomas2015protocol}  and
Atomic-swap~\cite{atomic-swap}
for inter-blockchain transactions;
 these decentralized PBT networks are promising for
addressing scalability, efficiency, and interoperability concerns with blockchains 
through off-chain transactions requiring no expensive mining efforts.
In fact, at a recent blockchain event, the InterLedger team demonstrated a transaction through seven 
blockchains including those in Bitcoin, 
Ethereum, and Ripple~\cite{sevenledgers}.

Unlike in blockchain-based PBT networks such as Ripple or Stellar,  
two users $u$ and $v$ in a decentralized PBT network \emph{locally} maintain a weighted link
(also called a payment channel, 
state channel, or credit link, depending on the application).
The link's weight characterizes the amount of funds (or assets) that one user can transfer to the other,
the exact nature of the link depending on the application.
For instance, in a credit network, the weight defines the difference between 
the amount of credit $u$ is willing to grant $v$ and the amount $v$ already
owes $u$.

A PBT network builds on top of three key algorithms: \emph{routing},  
\emph{payment} and \emph{accountability}. The routing algorithm is in charge of finding paths with 
enough funds from sender to receiver. The payment algorithm settles the funds between 
sender and receiver along the paths connecting them. Finally, the accountability algorithm 
allows the resolution of disputes in the presence of misbehaving users. 

While frequently omitted or disregarded as 
an orthogonal problem, the design of the routing algorithm is key to the PBT network's effectiveness, 
characterized by the fraction of successfully resolved transactions; efficiency, characterized by the delays experienced during a transaction as well as the overhead created by transactions; and scalability, 
 characterized  by the ability of a PBT network to maintain effectiveness and efficiency for 
 a growing base of users and transactions. 
 
Whereas industry supposedly considers efficiency, effectiveness, and scalability to be the main concerns for designing a routing algorithm, we additionally emphasize the need for privacy. 
Otherwise, the routing algorithm might reveal sensitive information such as the transaction value, the identity of sender and receiver, and the debt of one user to another.  
In this paper, we stress that all of effectiveness, efficiency, scalability, and privacy
are important to the design of a routing algorithm. 
A routing algorithm lacking any of these key properties is unlikely to be deployed.

The few routing algorithms proposed so far for PBT networks fail to achieve either 
privacy, efficiency, or scalability. For instance, the routing algorithm in Ripple and Stellar 
relies on a public blockchain that logs the complete PBT network, 
thereby introducing blockchain fees and impeding privacy. 
Canal~\cite{viswanath2012canal} relies on a single server to store the complete PBT network, 
 find paths, and settle payments between users. Therefore, the server is 
trivially aware of all links between users and their transactions. 
PrivPay~\cite{moreno15privpay} leverages trusted hardware to encrypt the PBT 
network data at the server and uses oblivious algorithms to hide the access patterns, thereby 
increasing the privacy for the links between users and their payments. Nevertheless, PrivPay 
 still suffers from a single point of failure and low scalability. Flare~\cite{prihodko2016flare}, 
 a routing algorithm for the Lightning Network, requires every user in the path from sender to 
 receiver to send the current fund amounts for their payment channels to the sender, thereby leaking 
 sensitive information~\cite{prihodko2016flare}.
The most promising approach with regard to privacy is SilentWhispers~\cite{malavolta17silent}, 
a decentralized PBT network without a public ledger. However, as we show in this paper, 
the routing algorithm in SilentWhispers lacks efficiency.

In this work, we present SpeedyMurmurs, a routing algorithm for PBT networks 
that provides formal privacy guarantees in a fully distributed setting and outperforms the state-of-the-art 
routing algorithms in terms of effectiveness and efficiency. 
SpeedyMurmurs extends VOUTE~\cite{roos2016anonymous}, a privacy-preserving embedding-based~\cite{papadimitriou2004conjecture} routing algorithm for message delivery in route-restricted P2P networks.  
Targeting message transmission in undirected and unweighted networks rather than payments, VOUTE 
is unequipped for dealing with weighted links and specifically changes of these weights as a result of previous transfers.  
\oursys combines the underlying ideas of VOUTE with the specifics of credit networks.      
In particular:
\begin{itemize}
\item SpeedyMurmurs considers both the available funds and the closeness to the destination of a neighbor when routing a payment, resulting in an efficient algorithm with flexible path selection.  
\item SpeedyMurmurs employs an on-demand efficient stabilization algorithm that reacts to changes of links if necessary but keeps the overhead corresponding to these changes low.
\item SpeedyMurmurs provides an improved handling of concurrent transactions by allowing nodes to proactively allocate exactly the amount of funds required for a transaction rather than barring concurrent transactions from using a link altogether or risking failures during the subsequent payment phase.
\item In our simulation study, which models a credit network and transactions based on a real-world dataset of Ripple ranging from 2013 to 2016, \oursys performs transactions at about twice the speed of \cnsysname and reduces the communication overhead of transactions by at least a factor of two while maintaining a similar or higher effectiveness. 
\item SpeedyMurmurs reduces the overhead of managing link changes by 2--3 orders of magnitude except for rare phases (approximately one per year) in the Ripple dataset corresponding to sudden rapid growth.
\item SpeedyMurmurs achieves \emph{value privacy}, i.e., the value of a transaction remains hidden, as well as \emph{sender} and \emph{receiver privacy}, i.e., the identities of the two users remain hidden from the adversary.   
\end{itemize}
In summary, SpeedyMurmurs offers an effective, efficient, and scalable solution for privacy-preserving routing in PBT networks, thus being a promising candidate for upcoming deployment of such networks.  Our release of the initial results initiated a discussion about the deployment of SpeedyMurmurs or related algorithms in the context of the Lightning network.\footnote{\url{https://lists.linuxfoundation.org/pipermail/lightning-dev/2017-November/000798.html}}


\section{State of the Art and Limitations}
\label{sec:background}

We first briefly overview the notion of a PBT network. 
Then, we introduce the concepts of landmark routing and embedding-based routing, including the description 
of \cnsysname~\cite{malavolta17silent}, a PBT network based on landmark routing, and VOUTE~\cite{roos2016anonymous}, an embedding-based routing algorithm, which we adapt to PBT networks in Section \ref{sec:algo}.

\subsection{PBT Networks}
In a PBT network, pairs of \emph{\users} locally maintain links weighted with 
application-dependent \emph{\money}. In the Lightning Network for instance, 
two \users create a link by adding a \emph{deposit} transaction in the blockchain 
and update such links by locally adjusting their deposit's value. 
The Lightning Network thereby reduces the load on the blockchain and 
it has become the most promising alternative for scaling Bitcoin. 

The payment operation in a PBT network involves a path of intermediate users 
who adjust their links pairwise to effectively settle \money between a \sender and a \receiver. 
In the Lightning Network, each intermediate \user increases her deposit's value
with 
their predecessor on the path by the transaction amount. 
Similarly, she decreases the deposit's value 
with her successor by the same amount. However, 
a payment cannot be performed without a \emph{routing} algorithm to find the path itself at first.

\subsection{Landmark Routing}

The landmark routing technique~\cite{tsuchiya1988landmark} enables the computation of 
a subset of paths between a sender and a receiver in a PBT network without relying
on the cost-intensive max-flow approach.
The key idea of landmark routing is to determine a path from sender to receiver through 
an intermediate node, called a \emph{landmark}, usually a well-known node of high connectivity. 
Using several such landmarks increases the number of computed paths between sender and receiver.
While landmark routing does not discover all possible paths and hence might lead to a lower
probability for a successful payment, past work indicates that the decrease of success is small
in comparison to the gain in performance~\cite{viswanath2012canal,moreno15privpay}.

Initially, each landmark starts two instances of the 
Breadth-First Search (BFS) algorithm, resulting in two spanning trees. In the first instance, only forward edges are considered and 
shortest paths from the landmark to each node are calculated. The second instance 
considers only reverse edges and results in shortest paths between each node and the landmark. 
As PBT networks change over time, landmarks repeat this initialization process periodically.

The path discovery between a sender and receiver then concatenates the path from the sender to the landmark
(using reverse edges) and the path from the landmark to the receiver (using forward edges).   
The sender can send \money along the path as long as the amount of \money is at most as high as the available
credit on each link on the path.

There are two versions of landmark routing. The first version, which we call \emph{landmark-centered}, always concatenates a path from the
source to a landmark and from the landmark to the destination.
The second version, which we call \emph{tree-only routing},
discovers the shortest path in the BFS tree, which does not necessarily contain a landmark.

\begin{figure*}[t]
\centering
\includegraphics[width=0.7\linewidth]{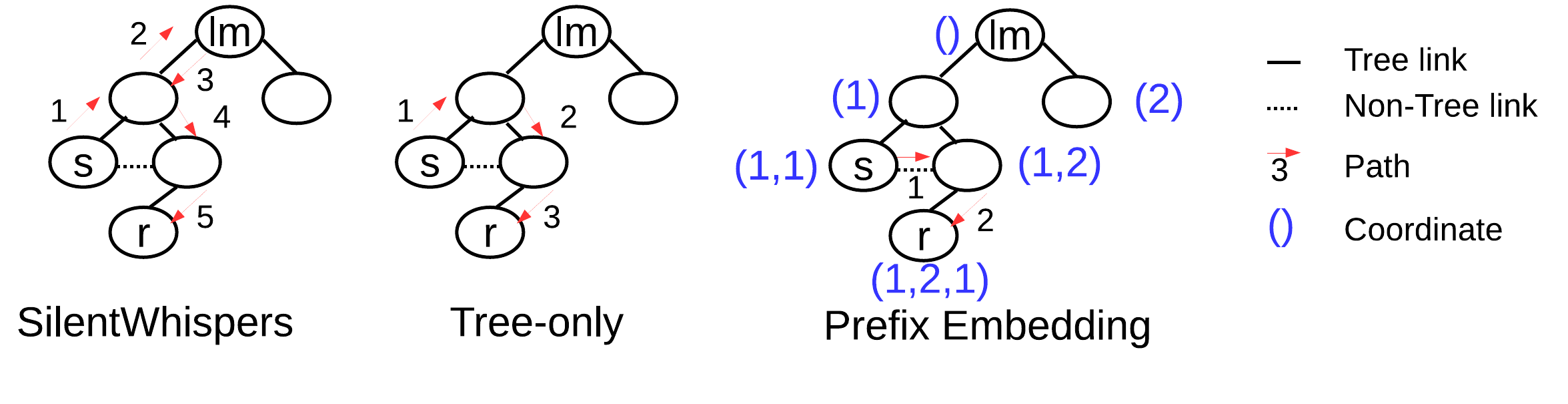}

\caption{Examples of different spanning tree routing schemes for landmark $lm$, sender $s$, receiver $r$.}
\label{fig:concepts-no-infer}
\vspace{-1em}
\end{figure*}

\subsubsection{Landmark Routing in SilentWhispers}
\cnsysname utilizes landmark-centered routing to discover multiple paths and then performs multi-party
computation to determine the amount of \money to send along each path. 
The initialization of the landmark routing follows the above description, using multiple landmarks
that perform periodic BFSs. 

The actual payment relies on two operations: a probe operation and the actual payment operation.
We here describe the probe operation as it performs the routing and decides on the credit to be transferred along each path.
The payment operation then merely executes the transfers suggested by the probe operation in a secure manner.  

At the core of the probe operation is a secret-sharing-based multiparty computation 
that computes the credit available in a path. 
After discovering paths between sender and receiver using landmark routing, 
each pair of adjacent users in the path 
sends a share of their link's value to each of the landmarks. The sender and receiver must  
construct additional shares that act as padding in order 
to hide the actual length of the path, and effectively preserve the identities of 
the actual sender and receiver. 
With the help of cryptographic signatures, relying on fresh keys to hide the identities of the nodes on the path,
and using multiparty computation, the landmarks determine shares that the sender can combine to obtain the minimal available credit $z_i$ of the \ith\ path. 
If the sum of the $z_i$ values 
is at least equal to the total payment amount, the sender assigns values $c_1, \ldots , c_{\numT}$ to the paths such that $c_i \leq z_i$. 
The result of the probe operation are these values $c_i$ and handles to
the paths,
which the payment operation leverages to perform the actual transfer.

\subsubsection{Weaknesses of SilentWhispers}
Based on the above description, we identify four issues related to the implementation of 
the routing algorithm in SilentWhispers. 
First, the periodic tree creation (execution of BFS) 
fails to take into account changes in the network immediately, 
which can lead to significant failure rates due to outdated information. 
 Moreover, periodic tree creation 
 induces unnecessary overhead due to re-computation for parts of the PBT network that might not have changed.  

Second, as SilentWhispers relies on landmark-centered routing, all paths include the landmarks even if i) the 
sender and receiver of a payment are in the same branch, or 
ii) there is a short path between sender and receiver but the links are not part of the spanning tree.  
Thus, the overall path used for the payment 
can be unnecessarily long, leading to longer delays 
and a lower success ratio due to the increased chance of encountering 
at least one link without enough funds. 

Third, the probe operation requires that the nodes  included in a transaction path send 
shares to all landmarks. This means that the transaction overhead scales 
quadratically in the number of landmarks. 

Fourth, \cnsysname does not provide a suitable solution for concurrency.
Assume that one or more probe operations aim to use the same link. 
The probe operation can either provide the same amount of available credit for both links or block use
of the link for some time after the first time a probe operation finds a path containing the link.
The former can lead to failures later on as the available credit, while sufficient for each transfer on its own, might not cover multiple transfers.  
While a block prevents such complications, it increases the likelihood of failures because probe operations cannot use certain links, which might have enough credit to execute multiple transactions.   
 Hence, both approaches to concurrency have severe drawbacks. 
 
In summary, landmark routing as used in \cnsysname has various weaknesses that we aim to overcome in this work.

\subsection{Embedding-based Routing}
\label{sec:embedding}

Peer-to-peer PBT networks differ from common peer-to-peer networks as the connections between peers are predefined and cannot be changed to improve the quality of the routing. Due to their fixed structure, peer-to-peer PBT networks are route-restricted and hence are closely related to Friend-to-friend (F2F) networks, which restrict connections to peers sharing a mutual trust relationship. As a consequence, we summarize the state-of-the-art approach to routing in F2F networks, namely embedding-based routing~\cite{papadimitriou2004conjecture,roos2016anonymous}.

Embeddings rely on assigning coordinates to nodes in a network and having nodes forward packets based on the distances between coordinates known to that node and a destination coordinate.
Greedy embeddings are similar to landmark routing in that they assign coordinates based on a node's position in a spanning tree. However, greedy embeddings disregard the spanning tree  after assigning the coordinates and in particular discover shorter paths using links that are not in the spanning tree. We refer to links that are not contained in the tree but are used during routing as \emph{shortcuts}. When a node $v$ forwards a message  addressed to a destination coordinate,
$v$ chooses the neighbor with the coordinate closest to the destination coordinate to forward the message to. Hence,  $v$ might either use a link in the spanning tree (forwarding to a child or parent), or a shortcut. 

Despite the fact that routes can contain shortcuts, there is no guarantee that routes with shortcuts exist. Hence, the links in the trees provide the guarantee that the routing works and removing any such links likely leads to failures. In the absence of shortcuts, embedding-based routing is identical to tree-only routing. As a consequence, it is important to adapt the tree when the nodes or links change.

Prefix Embedding~\cite{hofer2013greedy} is a greedy embedding that enables routing of messages in F2F overlays. 
As illustrated in \cref{fig:concepts-no-infer}, 
Prefix Embedding assigns coordinates in the form of vectors, starting with an empty vector at the landmark/root. 
Each internal node of the spanning tree enumerates its children and appends the enumeration index of a child to its coordinate to obtain the child coordinate.  
The distance between two such coordinates corresponds to the length of the shortest path in the spanning tree between them; i.e., the distance of two coordinates $\id(u)$ and $\id(v)$ with $|\id(w)|$ denoting the coordinate length of node $w$ and  $\var{cpl}(\id(u),\id(v))$ denoting the common prefix length is 
\begin{align}
\label{eq:dist}
d(\id(u),\id(v)) = |\id(u)|+|\id(v)|-2\var{cpl}(\id(u), \id(v)).
\end{align}   
Based on Eq.~\ref{eq:dist}, nodes determine which neighbor is closest to the receiver in terms of their coordinates' distance and forwards a message accordingly.  
\cref{fig:concepts-no-infer} displays an example to illustrate the difference between various tree-based routing schemes and illustrates the coordinate assignment in Prefix Embedding.

\subsubsection{Prefix Embeddings in VOUTE}
VOUTE~\cite{roos2016anonymous} is a routing algorithm building upon Prefix Embedding with the goal of anonymous and efficient message delivery for a dynamic route-restricted network; i.e., a network that does not allow the establishment of links between arbitrary nodes.
We quickly describe how VOUTE addresses the issues of privacy and dynamics. 

Prefix Embedding reveals the unique coordinate of the receiver. In contrast, VOUTE allows nodes to provide 
\emph{anonymous return addresses} instead of their original coordinates. A receiver generates a return address by padding its coordinate to a fixed length and generating keyed hashes of the coordinate's elements. The anonymous return address is then 
composed of the tuple (keyed hashes, key), where the key allows forwarding nodes to determine the common prefix length required in Eq.~\ref{eq:dist}.  
Based on the common prefix length of the receiver's coordinate $\id(r)$ and a neighbor's coordinate $\id(u)$, forwarding nodes can compute $d(\id(u), \id(r))+\Delta$ with $\Delta$ corresponding to the constant length of coordinates after the padding.
Hence, they can forward the message along the same path as when using clear-text coordinates, while maintaining the privacy of the receiver's true coordinates.

The original Prefix Embedding coordinates reflect an enumeration and hence have little entropy. As a consequence, VOUTE replaces the enumeration index with random $b$-bit numbers; e.g., for $b=128$. In this manner, guessing the coordinate of an unknown node becomes computationally unfeasible for an adversary.    

Rather than periodically reconstructing the spanning tree, VOUTE addresses dynamics with an on-demand stabilization algorithm. 
When constructing the tree, nodes send invitations to all neighbors stating their coordinate and offering to become a parent. 
Each node accepts one such invitation but keeps the most recent invitation of all neighbors to quickly react to network dynamics. 
If nodes establish a new link, nodes already contained in the spanning tree offer invitations to their new neighbors. If a node is not yet part of the tree, it accepts the invitation. Otherwise, it stores it for future consideration.
On the other hand, if a link in the spanning tree ceases to exist, the child node and all its descendants choose a new parent based on their remaining invitations. They then disseminate their new coordinate to all neighbors.
In this manner, spanning trees and embeddings have an on-demand repair mechanism rather than periodic full re-computation as in landmark routing.

\subsubsection{Limitations of VOUTE}
VOUTE has not been defined in the context of PBT networks and therefore 
presents several limitations that must be overcome before considering 
it as a routing algorithm in PBT networks.
In particular, VOUTE has incompatible assumptions with regard to the nature of links and topology dynamics in a PBT network.  

First, VOUTE considers \emph{undirected} and \emph{unweighted} links between pairs of users. In a PBT network instead, 
links are directed and weighted, as are payments. 
While all links allow message transfer in VOUTE, a link in a PBT network might not hold enough \money
to perform a payment. 
The directed nature of the links indicates that VOUTE's construction algorithm is insufficient as it is unclear how to deal
with unidirectional links. If unidirectional links are part of the spanning tree, a node (and it descendants) might only be able to send
or receive \money but not both. 
The weighted nature of links and the impossibility to use links for all payments contradicts one of the key assumptions of VOUTE's algorithm, namely that in the absence of link failures, all links can transfer messages. Therefore, to apply VOUTE in the context of PBT networks,
it is necessary to design algorithms that deal with weighted links and transfers. 

Second, VOUTE considers dynamics in the form of nodes joining and leaving the network. However, in PBT networks, the weights of the links are the main source of change. In particular, each successful transaction might change several links. A variant of VOUTE for PBT networks would likely be inefficient if it reacts to all of these changes. Deciding on when and how to adapt to changes of links is important for the design of such a variant.   

Finally, VOUTE does not have to deal with concurrency issues. While concurrent message transfers might increase delay and congestion, they do not change the capacity of links and transmitting a message does not affect the ability of the link to transmit future messages. 
However, separated probe and payment operations as in \cnsysname , creates concurrency issues. \cnsysname provides insufficient solutions here, so we require a new concurrency algorithm.     

In summary, although VOUTE presents an interesting alternative to landmark routing as 
implemented in SilentWhispers for the routing operation, its application in PBT network  
scenarios is not straightforward. 

\section{System Model and Goals}
\label{sec:model}

We start with a generic system model for distributed routing algorithms, followed by our privacy goals and our performance metrics.

\subsection{Our Model}
\label{sec:system-model}

We model a distributed \paysys $(G,\weight)$ as a 
directed graph $G=(V,E)$ and a weight function $\weight$ on the set of edges.  
The set of nodes $V$ corresponds to the participants of the \paysys.
A link (edge) from node $u$ to $v$ exists if $u$ can transfer funds to $v$. 
We define the set of outgoing neighbors of a node $v$ as $N_{\textit{out}}(v)=\{u \in V: (v,u) \in E\}$. Correspondingly, 
we define the set of incoming neighbors of a node $v$ as $N_{\textit{in}}(v)=\{u \in V: (u,v) \in E\}$. 
Furthermore, a path $\pathvar$ is a sequence of links $e_1 \ldots e_\pathlength$ with $e_i=(v_{i}^1, v_{i}^2)$, 
and $v_{i}^2=v_{(i+1)}^1$ for $1\leq i \leq \pathlength-1$. 
Moreover we denote by $\lm=\{\lmid_1, \ldots , \lmid_{\numT}\}$ a set of nodes, called landmarks, 
that are well known to other users in the \paysys. We denote by $|\lm|$ the size of the 
set $\lm$.

The function $\weight$ describes the amount of \money that can be transferred between two nodes 
sharing an edge. We thereby abstract from the specific implementation of the function $\weight$.  
For instance, in the Bitcoin Lightning Network, the function $\weight: E \rightarrow \mathbb{R}$ defines the number of bitcoins
$u$ can  transfer to $v$ in a payment channel opened between them. 

We define the \money available in a path $e_1, \ldots, e_\pathlength$ as the \emph{minimum} $\weight(e_i)$. 
Moreover, we define the the net balance of a node $v$ as 
$\displaystyle{\cnode(v)=\sum_{u \in N_{\textit{in}}(v)} \weight(u,v) - \sum_{u \in N_{\textit{out}}(v)} \weight(v,u)}$.

\subsubsection{Operations}
\label{sec:paysys-operations}
Routing  in a \paysys consists of a tuple of algorithms (\setRoute, \setC, \routePay) defined as follows: 
\begin{asparadesc}
\item $\setRoute(\lm)$: Given the set 
$\lm=\{\lmid_1, \ldots, \lmid_{\numT}\}$ of landmarks, $\setRoute$ initializes the routing information required by 
each node in the \paysys. 

\item $\setC(c,u,v)$: Given the value $c$ and the nodes $u$ and $v$, $\setC$ sets $\weight(u,v) = c$. In addition, $\setC$ might alter the routing information initially generated by \setRoute.

\item $((\pathvar_1, c_1), \ldots, (\pathvar_{|\lm|}, c_{|\lm|})) \gets \routePay(c,u,v)$. Given a value $c$, a sender $u$ and a receiver $v$, $\routePay$ returns a set of 
tuples $(\pathvar_i, c_i)$, denoting that $c_i$ \money are routed through the path 
described by $\pathvar_i$. 
\end{asparadesc}

\paragraph{Correctness}
A key property of a PBT network is correctness. Intuitively, correctness indicates that the routing algorithm i) suggests to spend  the desired \money $c$ rather than a higher value and ii) suggests paths that indeed have sufficient \money.  
Let $(\setRoute, \setC, \routePay)$ be the routing operations 
of a \paysys and let $\pathlength_i$ denote the length of the \ith\ discovered path. 
We say that the \paysys is \emph{correct} if for all results 
 $((\pathvar_1, c_1), \ldots, (\pathvar_{|\lm|}, c_{|\lm|}))$ of  
$\routePay(c,u,v)$, the following two conditions hold: 
\begin{itemize}
\item $\sum_{i} c_i \leq c$
\item For each $\pathvar_i := e_{i}^1, \ldots, e^{\pathlength_i}_i$ and each $e^{j}_i$, 
$c_i \leq \weight(e^{j}_i)$. 
\end{itemize}

We note that the $\routePay$ operation could return paths that contribute 
$\sum_i c_i < c$, and it is still considered correct. This accounts for the cases 
where the \paysys does not provide enough liquidity between the 
sender and receiver to perform a transaction.

\subsection{Attacker Model}
\label{sec:attacker-model}

We consider a fully distributed network.
Our primary attack scenario is companies and individuals interested in a user's
financial situation rather than governmental security agencies.
 The adversary controls a subset of the nodes in the network
either by inserting its own nodes or corrupting existing nodes. 
We assume that the adversary cannot choose the set of users at will, as some users will
be harder to corrupt by social engineering or malware attacks.  
In general, we assume that the attacker does not know \emph{all} links and nodes in the network and
in particular cannot access routing information locally stored at non-compromised nodes. 
The assumption that the attacker does not know the complete topology of a large-scale distributed
system with participants from a multitude of regions and countries seems realistic for our attack
scenario. If we indeed have an adversary that knows the full topology, we might not be able to hide
the identities of sender and receiver but can still hide the transaction value. 
 
Our adversary aims to undermine the privacy rather than perform a large-scale denial-of-service attack.
We argue that the primary defense against denial-of-service attacks is detection and expulsion of
malicious nodes. While related to routing, different operations are required
for realizing detection and expulsion, and they are out of scope for this paper. 

 While our overall adversary model limits the adversary's capacities, 
 we nonetheless define our value privacy goals for an attacker that has a global view of the topology, 
 indicating that we can still achieve some privacy against a stronger adversary. 

\subsection{Privacy Goals}
\label{sec:privacy-goals}
The hope that cryptography and decentralization might ensure robust privacy 
was among the strongest drivers of Bitcoin's and blockchains' early success. 
We expect businesses and customers employing the PBT networks to be interested 
in hiding their transactions from competitors and even service providers. 
Therefore, ensuring privacy for path-based transactions is important.

Like PrivPay~\cite{moreno15privpay}, SilentWhispers~\cite{malavolta17silent}, Fulgor~\cite{malavolta17PCN}, and Rayo~\cite{malavolta17PCN}, we aim to hide values (\textit{value privacy}), 
and the identities of sender and receiver (\emph{sender/receiver privacy}) of
path-based transactions.  We use the term \emph{transaction privacy} to
refer to meeting all three of these notions.
Next, we informally describe these privacy properties for PBT networks, and refer the readers to the PrivPay paper~\cite{moreno15privpay}
for the formalized versions defined in the context of credit networks.

\paragraph*{Value Privacy} 
A PBT network achieves value privacy if
it is not possible for any adversary to determine the total
value of a transaction between non-compromised users
as long as none of the employed intermediate nodes is compromised.

Let $s$ and $r$ be two non-compromised users, and let $(\pathvar_1, c_1),
\ldots, (\pathvar_{|\lm|}, c_{|\lm|})$ be the result of a $\routePay(c,s,r)$ operation. 
If for every path $\pathvar_i$, 
all nodes on that path are non-compromised, 
the adversary (even a global passive adversary) 
obtains no information about the transaction value $c$. 

Notice that, as elaborated in Section~\ref{sec:privacy},
we can provide a weaker form of value privacy even when the adversary compromises some intermediate nodes
as long as all nodes on at least one of the employed paths remain non-compromised.

\paragraph*{Sender Privacy} 
A PBT network achieves sender privacy if
it is not possible for any adversary to determine the sender 
in a path-based transaction between non-compromised users.

In particular, for two non-compromised users $s$ and $r$, 
the attacker should not be able to determine the sender $s$ of any routing operation $\routePay(z,s,r)$,
unless she has complete knowledge of $s$'s incoming links 
i.e., she knows the set $N_{\textit{in}}(s)$ though not necessarily the funds of the links  $e \in N_{\textit{in}}(s)$.

Note that although the local attacker without a global view of the network might know (and even control) all nodes in $N_{\textit{in}}(s)$,
she might not be aware that she does control {\em all} such nodes. As a consequence, similar to P2P anonymity systems \cite{Crowds,AP3,ShadowWalker}, 
controlling all neighbors does not automatically mean she can be sure that 
$s$ did initiate the routing.
Therefore, we expect the sender privacy to hold even when the attacker controls all nodes in $N_{\textit{in}}(s)$ 
for the sender $s$ but does not know that she does control the whole set.

\textit{Receiver Privacy} is defined analogously to sender privacy, and the adversarial assumptions also remain the same except that instead of 
neighboring nodes  $N_{\textit{in}}(s)$  of the sender $s$, now for receiver privacy, we  consider  $N_{\textit{out}}(r)$  of the receiver $r$.

\subsection{Performance Metrics}
In this section, we describe the performance goals to be achieved by a routing algorithm, which we 
denote generically by $\TA$. In the following, 
we denote by $(G_t, \weight_t)$ the snapshot of a \paysys at time $t$. Note that although 
we abstract away the payment and accountability algorithms in this work, a \paysys must 
implement them and therefore a \paysys is dynamic. Let $\{(t_i,c_i,s_i,r_i)\}$
be a set of payment requests from $s_i$ to $r_i$ for an amount $c_i$ at time $t_i$.

The performance of a routing algorithm $\TA$ is characterized by the following four metrics:
\begin{itemize}
\item \textbf{Success ratio}: 
Let $((\pathvar_1, c_1), \ldots,$ $(\pathvar_{|\lm|}, c_{|\lm|}))$ be the set of paths 
returned by $\routePay(c, s, r)$ as implemented in $\TA$. We consider the transaction successful 
only if $\sum_i c_i = c$.  The success ratio describes the fraction of transactions  
that are successful.\footnote{This inherently assumes a payment algorithm that always succeeds 
after a route with enough credit has been found. We thereby abstract away the details of the 
payment algorithm.} 
 
\item \textbf{(Hop) Delay}: The delay of $\TA$ with regard to a transaction $(t_i,c_i,s_i,r_i)$ is the difference between the time of termination and the initiation time $t_i$. In the absence of a concrete implementation including realistic computation and communication latencies, we provide an abstract measurement of the delay as follows. Let $m_1$ and $m_2$ be messages sent by $\TA$. We say $m_2$ is \emph{subsequent} to $m_1$ if a node sends $m_2$ as a result of receiving $m_1$. The hop delay is the length of the longest chain of subsequent messages sent by $\TA$.
\item \textbf{Transaction Overhead}: Nodes exchange messages to execute a transaction $(t_i,c_i,s_i,r_i)$. The transaction overhead denotes the number of exchanged bits. 
As before, in the absence of a concrete implementation, we abstract it by assuming equal-sized messages for each implementation and stating the number of messages as the overhead. 
\item \textbf{Stabilization Overhead}: Analogously to the transaction overhead, the stabilization overhead corresponds to the number of bits sent within a certain time interval to maintain necessary state information, 
as required by the implementation of the operation $\setRoute()$ in $\TA$. Again, we abstract from the concrete implementation by stating the number of messages instead of the number of bits. 
\end{itemize}
The first two metrics heavily impact the perceived quality of service while the latter two directly relate to network congestion and hence impact the delay. Furthermore, the overhead determines the load on the user devices.



\section{Our Construction}
\label{sec:algo}

In this section, we first describe the key ideas of our design and then detail  the three operations in our routing algorithm. We present pseudocode for centralized versions of the algorithms, which allows presenting the algorithms in a linear compact manner. We then describe how the distributed versions used within a PBT network differ from the centralized version. 

\subsection{Assumptions}  
Each user in the \paysys locally maintains the information of the links with her neighbors. We 
further assume that users sharing a link can send messages to each other through an 
authenticated and confidential communication channel. Moreover, we assume that there exist 
a set of nodes, called landmarks, that are well known to other users in the 
\paysys. We note that these assumptions are in tune with other distributed \paysyss such as 
SilentWhispers.

Throughout this section, we refer to links that have non-zero funds in both directions, i.e., links $(u,v)$ and $(v,u)$
with $\weight(u,v)>0$ and $\weight(v,u)>0$ as bidirectional. $u$ and $v$ have a unidirectional link if one of the two links does not exist or has zero funds. 

\subsection{Overview and Key Ideas}

We here describe the key ideas of \oursys with regard to the operations \setRoute, \setC, and \routePay.
In particular, we focus on the predominant differences to existing algorithms. 

\paragraph*{\setRoute}  
In this algorithm, we construct multiple embeddings, one for each landmark. 
As mentioned in Section~\ref{sec:embedding}, VOUTE offers an algorithm for BFS-based coordinate assignment that assumes unweighted and undirected links. We modify this algorithm by dividing it into two phases.
First, we only add bidirectional links. In the second phase of the algorithm, nodes that are not yet part of the spanning tree join by adding unidirectional links to the spanning tree.

\paragraph*{\setC} We first change the weight of the link and then adapt the embedding accordingly. 
VOUTE does not provide any guidance on how to react to changes of weights. In agreement with \setRoute , we decided to only initiate changes if the two nodes i) established a new link of non-zero weight (or set the value of link from 0 to a non-zero value), and ii) removed a non-zero link (or set its value to 0). 
If they established a new link, one of the nodes can choose the other as a parent if it does not have a parent or the link to its current parent only has credit in one direction. In contrast, if a link is removed, one of the nodes has to select a new parent (and coordinate) if the other node previously was its parent. Furthermore, any descendants of the affected node have to change coordinates.

\paragraph*{\routePay} The routing consists of three steps: i) the receiver generates anonymous return addresses and sends them to the sender,
ii) the sender randomly splits the transaction value on $|\lmid|$ paths, one for each landmark, and iii)
VOUTE's routing algorithm finds a path between sender and receiver, restricted to links that have sufficient funds.  
Our algorithm allows a flexible choice of routes, preferring paths with high funds. Determining the \money transferred along each path beforehand enables nodes to block a certain amount of credit during the probe operation and ensure that the subsequent payment succeeds without blocking all of the link's funds.

\subsection{Detailed Description}

\paragraph*{\setRoute}  
In the initialization phase, described in Algorithm~\ref{algo:embedCen}, we derive the embeddings. 
Iterating over all landmarks, Algorithm~\ref{algo:embedCen} assigns the landmark coordinate to be the empty vector (Line~\ref{algo:initEmpty}) and appends the landmark to a  queue (Line~\ref{algo:queueL}). The main loop of the algorithm then processes the queue. In each step, the algorithm removes a node from the queue (Line~\ref{algo:removeHead}) and considers all its neighbors. If a neighbor $n$ does not have a coordinate yet and is eligible to have one, the algorithm integrates $n$ into the spanning tree, assigns a coordinate by concatenating the parent coordinate and a random $b$-bit number, and appends it to the queue (Lines~\ref{algo:parent}-\ref{algo:queueN}).   
The criterion determining the eligibility to be part of the spanning depends on the phase of the algorithm: Initially ($\var{bi}=\var{true}$, Line~\ref{algo:bi}), a node is eligible if the available credit on the link to and from its potential parent is non-zero. In the second phase of the algorithm ($\var{bi}=\var{false}$), triggered by an empty queue (Lines~\ref{algo:uni}--\ref{algo:endif}),  all nodes can join the spanning tree. Note that Algorithm~\ref{algo:embedCen} does not prevent a child from choosing a parent such that they do not have funds in either direction. As such links do not serve any purpose in a PBT network, we assume that the network does not allow them. Alternatively, Algorithm~\ref{algo:embedCen} can check if the funds on a unidirectional link is non-zero before allowing a parent-child relation.  
The algorithm terminates once the queue is empty, indicating that all nodes in a connected graph have coordinates. 

\begin{algorithm}[t]
\caption{\setRoute}
\label{algo:embedCen}
\begin{algorithmic}[1]
\STATE \# Input: Graph $G$, landmarks $\lmid_1, \ldots ,\lmid_{\numT}$
\FOR{$i=1$ \TO $\numT$ } 
\STATE $\id_i(\lmid_i)=()$ \label{algo:initEmpty}
\STATE $q=$ empty queue
\STATE add $\lmid_i$ to $q$ \label{algo:queueL}
\STATE $\var{bi}=\var{true}$ \# first consider nodes with links in both directions  \label{algo:bi}
\WHILE{$q$ is not empty}
\STATE $\var{node} =$ remove head of $q$ \label{algo:removeHead}
\FORALL{$n$: neighbors of node}
\STATE $n$ stores $\id_i(\var{node})$
\STATE \# Assign coordinate if $n$ does not have one  
\IF{$\id_i(n)$ is not set}
\IF{($\weight(\var{node},n)>0$ \AND $\weight(n,\var{node})>0$) \OR
$!\var{bi}$}
\STATE $\var{parent}_i(n)=node$ \label{algo:parent}
\STATE $n$ chooses random $b$-bit number $r$
\STATE $\id_i(n)=\var{concatenate}(\id_i(\var{node}), r)$ 
\STATE add $n$ to $q$ \label{algo:queueN}
\ENDIF
\ENDIF
\ENDFOR
\STATE \# Add nodes with unidirectional links \label{algo:uni} 
\IF{$q$ is empty \AND $\var{bi}$}
\STATE $\var{bi}=\var{false}$
\STATE add all nodes $n$ with $\id_i(n)$ set to $q$ 
\ENDIF \label{algo:endif} 
\ENDWHILE
\ENDFOR
\end{algorithmic}
\end{algorithm}

In a distributed scenario, there are no central queues. Rather, nodes send messages to their neighbors when they join a spanning tree. Each message includes the index $i$ of the tree and the coordinate $\id_i(\var{node})$ of the potential parent.
Starting the second phase is tricky in a distributed scenario and will not be simultaneous for all nodes. Rather, we choose a time limit $\tau$ that represents an upper limit on the time the initialization should take. If a node $n$ receives a message of the form  $(i,\id_i(\var{node}))$ from a neighbor $node$ with only one link of non-zero weight, $n$ waits for time $\tau$. If none of $n$'s neighbors with bidirectional links to $n$ indicate that they are potential parents until the waiting period ends, $n$ selects $node$ as its parent. 

\paragraph*{\setC}

\setC\ reacts to a pair of nodes $(u,v)$ that want to change the value of their shared link to $c$. 
Algorithm~\ref{algo:setC} first determines if the value change should lead to coordinate changes.
In total there are three situations that indicate a need for a coordinate change:
\begin{enumerate}
\item New non-zero unidirectional link: One of the nodes is not yet part of the tree and should hence choose the other as their parent (Lines~\ref{algo:unsetSt}--\ref{algo:unsetEnd}) to be able to participate in the routing.
\item New non-zero bidirectional link: $u$ and $v$ share a bidirectional non-zero link and (without loss of generality) $u$ has only a unidirectional link to its current parent. Then $u$ should change its parent to $v$ if $v$ has a bidirectional link to its parent (Lines~\ref{algo:uniSt}--\ref{algo:uniEnd}). In this manner, a bidirectional connection replaces a unidirectional link in the spanning tree and increases the likelihood of successfully transferring \money. 
\item Removed link: $u$ is a child of $v$ or $v$ is a child of $u$ (Lines~\ref{algo:rmSt}--\ref{algo:rmEnd}). The child node should select a new parent to increase the number of non-zero links in the spanning tree and the likelihood of transferring \money. 
\end{enumerate}

\begin{algorithm}[t]
\caption{$\setC$}
\label{algo:setC}
\begin{algorithmic}[1]
\STATE \# Input: Graph $G$, $u,v \in V$, new value $c$
\STATE $\var{old} = \weight(u,v)$ \# Previous value of $\weight(u,v)$
\STATE \# check if coordinate change necessary 
\FOR{$i=1$ \TO $\numT$ }
\STATE $\var{reset}=\var{null}$ \# node whose coordinate should change 
\STATE \# case:add link
\IF{$old==0$ \AND $c>0$}
\STATE \# If one node does not have a coordinate \label{algo:unsetSt}
\IF{$\id_i(v)$ is not set and $\id_i(u)$ is set}
\STATE $\var{reset}=v$
\ENDIF 
\IF{$\id_i(u)$ is not set and $\id_i(v)$ is set}
\STATE $\var{reset}=u$
\ENDIF \label{algo:unsetEnd}
\STATE \# One node has unidirectional link to parent \label{algo:uniSt}
\IF{$\var{reset}=\var{null}$}
\IF{$\weight(u,v)>0$ \AND $\weight(v,u)>0$}
\STATE $a_1=$ $\big(\weight(u, \var{parent}_i(u))==0$ \OR\\ \hspace{2.5em} $\weight(\var{parent}_i(u),u)==0\big)$
\STATE $a_2=$ $\big(\weight(v, \var{parent}_i(v))==0$ \OR\\ \hspace{2.5em} $\weight(\var{parent}_i(v),v)==0\big)$
\IF{$a_1$ \AND $!a_2$}
\STATE $\var{reset}=v$
\ENDIF
\IF{$a_2$ \AND $!a_1$}
\STATE $\var{reset}=u$
\ENDIF
\ENDIF
\ENDIF \label{algo:uniEnd}
\ENDIF
\STATE \# case:remove link
\IF{$\var{old}>0$ \AND $c==0$}
\IF{$\var{parent}_i(u)==v$} \label{algo:rmSt}
\STATE $\var{reset}=u$
\ENDIF
\IF{$\var{parent}_i(v)==u$} 
\STATE $\var{reset}=v$
\ENDIF \label{algo:rmEnd}
\ENDIF
\STATE \# change coordinates
\IF{$\var{reset}$ != $\var{null}$}
\STATE delete coordinates of $\var{reset}$ and descendants
\STATE have nodes choose new parent
\ENDIF 
\ENDFOR
\end{algorithmic}
\end{algorithm}

If one of $u$ or $v$ changes its parent, all descendants remove their coordinates and inform their neighbors of the removal. 
Afterwards, they all choose a new parent and corresponding coordinate. In agreement with the initialization \setRoute , nodes first consider only neighbors to whom they have non-zero links in both directions. However, if a node does not have such links to any neighbor, it considers links in one direction. If they have several suitable parents, they choose their parent randomly from those candidates with the shortest coordinates, as having short routes to the landmark reduces the lengths of paths~\cite{roos2016anonymous}. 
After choosing a new coordinate, nodes forward the new coordinate and the tree index to all their neighbors. We do not present the pseudocode, as it is very similar to Algorithm~\ref{algo:embedCen}.

The distributed variant of Algorithm~\ref{algo:setC} follows the same principles but requires the exchange of messages for nodes to communicate information. $u$ and $v$ exchange information about the link to their parents. Each of them then individually decides if they want to add or remove the other as a parent. 
Starting from the node $\var{reset}$ that aims to reset its coordinate, all descendants inform their neighbors first that they remove their old coordinate for the tree $i$. Children of a node in turn remove their own coordinate and send the respective messages. In the second phase, nodes select their new coordinates and inform their neighbors. 
As the two phases are likely to run in parallel in the distributed setting, nodes have to ensure that they do not choose a previous descendant as a parent before the descendant chooses a new coordinate. However, the nature of the coordinates makes it easy to prevent such cycles in the tree by disallowing a node $v$ from choosing a parent whose coordinate contains $v$'s previous coordinate as a prefix. 

\paragraph*{\routePay}
\routePay\ discovers a set of paths from the sender to the receiver. It corresponds to the probe operation in \cnsysname. Algorithm~\ref{algo:route} divides the process into three steps: i) generation of receiver addresses (Lines~\ref{algo:addSt}--\ref{algo:addEnd}), ii) splitting the total transaction value $c$ randomly on $\numT$ paths, and iii) finding paths for all embeddings that can transmit the required value. 

First, the receiver generates anonymous return addresses $\var{add}_1, \ldots , \var{add}_{\numT}$ for all landmarks and sends them to the sender (Lines~\ref{algo:addSt}--\ref{algo:addEnd}).
Second, the sender splits the transaction value randomly between all paths (Line~\ref{algo:split}). By defining a per-path value before routing, we i) avoid the costly multiparty computation of \cnsysname and ii) allow the algorithm to choose between several possible routes. Avoiding the multiparty computation of the minimum also removes a privacy leakage, as knowing the minimum value of funds available on the complete path naturally reveals information about the individual links.

Third, the route discovery starts at $v$ and each node selects a neighbor to be the next node on the route. In VOUTE, each node would select the neighbor with the coordinate closest to the destination, using the function $\tilde{d}$ that compares a coordinate with an anonymous return address.  
 However, such a choice might not be suitable for routing \money as the link might have insufficient available credit. As a consequence, the routing only considers links $(v,u)$ with \emph{guaranteed} available credit $\weight_A(v,u)$ of at least $c_i$ (Line~\ref{algo:CSt}). We differentiate between available credit $\weight(v,u)$ and  \emph{guaranteed} available credit to deal with concurrency. $\weight_A(v,u)$ is a lower bound on the available credit if ongoing probe operations succeed.
Initially, $\weight_A$ equals the actual available credit $\weight$. We do not include the initialization in Algorithm~\ref{algo:route} as multiple concurrent executions of \routePay\ can impact $\weight_A(v,u)$ and the algorithm might start with $\weight_A(v,u)<\weight(v,u)$.  
If a probe operation indicates that a payment will transmit \money $c_i$ along a link $(v,u)$, 
we proactively decrease the \emph{guaranteed} available credit by $c_i$ (Line~\ref{algo:potA}) to keep future routings from using the link unless they require at most the guaranteed available credit. 
If the routing fails, we add $c_i$ to $\weight_A(v,u)$ again (Lines~\ref{algo:rePotSt}--\ref{algo:rePotEnd}). 
The routing fails if a node $v$ cannot find a neighbor with a coordinate closer to the destination than $v$'s coordinate and a link of sufficient guaranteed available credit.

\begin{algorithm}[t]
\caption{\routePay}
\label{algo:route}
\begin{algorithmic}[1]
\STATE \# Input: Graph $G$, sender $\var{src}$, receiver $\var{dst}$, value $c$
\STATE \# get addresses \label{algo:addSt}
\FOR{$i=1$ \TO $\numT$ } 
\STATE use VOUTE's algorithm to generate return address $\var{add}_i(\var{dst})$ 
\STATE $\var{dst}$ sends $\var{add}_i(\var{dst})$ to $\var{src}$
\ENDFOR \label{algo:addEnd}
\STATE \# value shares for each path 
\STATE $\var{src}$ splits $c$ into shares $c_1, \ldots , c_{\numT}$ \label{algo:split}
\STATE \# routing \label{algo:rSt}
\STATE $\var{path}_i =$ empty list of links 
\FOR{$i=1$ \TO $\numT$ } 
\STATE $v=\var{src}$
\STATE $\var{fail}=\var{false}$
\WHILE{$!\var{fail}$ \AND $v$ != $\var{dst}$}
\STATE $C=\{u \in N(v): \tilde{d}(\id_i(u), \var{add}_i(\var{dst}))< \tilde{d}(\id_i(u), \var{add}_i(\var{dst})), \weight_A(v,u)\geq c_i\}$ \label{algo:CSt}
\IF{$C$ not empty}
\STATE $\var{next}=$ $u$ in $C$ with minimal $\tilde{d}(\id_i(u), \var{add}_i(dst))$ \label{algo:CEnd}
\STATE $\weight_A(v,u)=\weight_A(v,u)-c_i$ \label{algo:potA}
\STATE $v=\var{next}$
\ELSE
\STATE $\var{fail}=\var{true}$ \# Routing failed
\ENDIF
\ENDWHILE \label{algo:rEnd}
\ENDFOR
\IF{routing failed} \label{algo:rePotSt}
\FORALL{$i=1\ldots \numT$, $e \in \var{path}_i$}
\STATE $\weight_A(e)=\weight_A(e)+c_i$
\ENDFOR
\ENDIF \label{algo:rePotEnd}
\end{algorithmic}
\end{algorithm}

Algorithm~\ref{algo:route} achieves correctness, as defined in Section~\ref{sec:model}, because i) $\sum_{i=1}^{\numT} c_i = c$ and ii) nodes always select links $e$ with $\weight(e)\geq\weight_A(e)\geq c_i$ on the \ith\ path.   

In the distributed variant of Algorithm~\ref{algo:route}, nodes send messages to the next node on the path, which contain the address $\var{add}_i$ and the partial value $c_i$. Nodes report failures and successes to the sender by sending messages along the reverse path. 
To account for messages getting lost, nodes also reset $\weight_A$ if a payment operation does not follow a probe operation within a certain time.

\subsection{Parameters}
Several parameters govern the performance of the above routing algorithm. First, the number $\numT$
of landmarks determines the number of returned paths. The transaction and stabilization overhead increases roughly linearly with $\numT$ as routing and stabilization is required for each landmark. Similarly, the delay corresponds to the longest route in any embedding and hence is likely to increase with $\numT$. 
The impact of $\numT$ on the success ratio highly depends on the scenario. 
The second parameter is $\numA$, the number of transaction attempts. 
A sender $s$ can attempt to perform a transaction up to $\numA$ times.
Only if all attempts fail, $s$ considers the transaction failed. 
$s$ chooses the interval between two consecutive attempts uniformly at random within an interval of length $\tl$.
A repeated transaction attempt executes the above routing algorithm for the same sender, receiver, and value but uses different shares $c_1, \ldots , c_{\numT}$. 
In addition to the parameters $\numT$, $\numA$, and $\tl$, the choice of the landmarks impacts the performance.
Commonly, landmarks are nodes corresponding to financial institutions and hence have a large number of links, possibly leading to spanning trees of a lower depth and a higher performance. 
We characterize the impact of these parameters in our performance evaluation.

\subsection{Privacy Analysis}
\label{sec:privacy}

Next, we argue that \oursys achieves the privacy goals 
proposed in~\cref{sec:privacy-goals}. 

\paragraph*{Value Privacy} Informally, we say that a \paysys achieves value privacy if 
the adversary cannot determine the value $c$ of a $\routePay(c,u,v)$ operation between 
two non-compromised users $u$ and $v$, if the adversary is not sitting in any of the 
involved routing paths. 

\oursys is a distributed \paysys and, in particular, the $\routePay$ is defined 
such that only users in the paths  
between the sender and receiver are involved. Therefore, if the adversary does not 
compromise any such users, she does not get any information about the 
routed value (because the point-to-point communications are encrypted) and thereby value privacy is achieved.

An alternative scenario appears when the adversary corrupts users 
in some of the paths between sender and receiver, but not all. 
In such a case, we cannot prevent the adversary from estimating $c$. As we have $c_i\geq 0$ for all $i=1\ldots \numT$, knowing a subset of these values naturally reveals information about the total value $c$, namely that $c \geq c_i$. Moreover, as  \oursys shares the value $c$ uniformly among the paths and uses 
only positive shares, an adversary can estimate $c$ as $\numT * c_i$ upon observing $c_i$.

\paragraph*{Sender Privacy} Informally, we say that a \paysys achieves sender privacy if an adversary 
cannot determine the sender $u$ in a $\routePay(c,u,v)$ operation. The adversary might compromise intermediate users
on the paths discovered by $\routePay(c,u,v)$ but does not compromise $u$ or $v$.

An attacker sitting on the path between sender $s$ and receiver $r$ might receive an anonymous 
routing address $add_i$ (e.g., the adversary managed to corrupt the sender's neighbor). 
Nevertheless, as \oursys is a distributed \paysys, 
the adversary cannot determine whether the actual sender is $s$ or 
another user $s'$ connected to $s$ through a direct link or a path of non-compromised users.
Sender privacy follows from the corresponding proofs for VOUTE~\cite{roos16voute}.

\paragraph*{Receiver Privacy} Informally, we say that a \paysys achieves receiver privacy if an adversary 
cannot determine the receiver $v$ in a $\routePay(c,u,v)$ operation. The adversary might compromise intermediate users
on the paths discovered by $\routePay(c,u,v)$ but does not compromise $u$ or $v$.   

As before, the adversary compromising the \user before the receiver $r$ might relay
to $r$  
an anonymous return address. Nevertheless, as shown in
the 
evaluation of VOUTE, 
an anonymous return address does not 
leak the corresponding user in the network. Therefore, 
the adversary cannot fully determine yet if 
$r$ is the actual receiver, or the routing message is intended for another receiver $r'$ connected 
to $r$ through a direct link or a path of non-compromised users.

\subsection{Summary}

In this section, we introduced \oursys , which proposes a privacy-preserving routing algorithm for PBT networks. Our key contributions in modifying VOUTE to the scenario of credit networks are i) the use of a two-phase construction algorithm to account for the existence of unidirectional links (Algorithm~\ref{algo:embedCen}), ii) the identification of criteria on when to apply on-demand maintenance (Algorithm~\ref{algo:setC}), iii) the design of a path discovery algorithm that can adaptively choose links based on both the available credit and the coordinates of the neighboring nodes and can handle concurrency (Algorithm~\ref{algo:route}).
Apart from using embedding-based routing, \oursys distinguishes itself from \cnsysname by splitting the credit between paths before the path discovery. In this manner, nodes can base their forwarding decisions on the amount of credit they should forward rather than only their neighbors' distances to the destination.  On the other hand, distributing funds before the path discovery prevents the algorithm from taking the overall available funds on the path into consideration. In the next section, we evaluate the impact of our design decisions on efficiency and effectiveness, analyzing in particular how the order of routing and fund distribution relates to the success ratio.


\section{Performance Evaluation}
\label{sec:simu}

In this section, we evaluate the performance of \oursys in comparison to the related work, in particular \cnsysname ' landmark-centered routing.

More precisely, we aim to answer the following research questions:
\begin{itemize}
\item How do \oursys and \cnsysname perform with regard to success ratio, delay, and overhead using a real-world dataset?
\item \oursys and \cnsysname differ in three major areas---routing algorithm, random credit assignment, and dynamic stabilization.  What is the impact of each of these modifications on the above performance criteria?
\item How do these results compare to the performance of other approaches? 
\item What is the impact of the landmark selection, the number of trees, and the number of transaction attempts?
\item How does the long-term evolution of the credit network affect the performance? 
\end{itemize}

We start by describing our simulation model and datasets. Afterwards, we specify the parameters of our simulation. 
Finally, we present and discuss our results. 

Generally, our simulation executes the routing algorithm and performs the payment (if successful). 
We include the payment to realistically assess the stabilization overhead due to link changes.  
However, we did not implement any security measures that are usually part of the payment because they do not affect the routing algorithm and its performance. 
In particular, we do not execute the link setup algorithm that ensures that neighboring nodes agree on the value of their link and later can settle disputes by providing signed statements of the changes. 

\subsection{Simulation Model}

We extended GTNA~\cite{schiller2013gtna}, a framework for graph analysis, to include our credit transaction mechanisms.
In particular, GTNA offers templates for routing algorithms and performance metrics.
We added functionality specific to PBT networks, in particular the functionality to dynamically update link weights.

Initially, our simulation constructs a credit network with nodes and links according to a provided description. 
Afterwards, we simulate a sequence of events in the credit network.
A list of transactions, changes to links, and periodic re-computations of the spanning tree (only required for \cnsysname), ordered
by their temporal occurrence, determined the sequence of events. 
In the absence of realistic latency and bandwidth models, we did not model concurrency in our simulation.
The simulation executed each event, including resulting changes to the spanning trees, before starting the next event.

We implemented two simulation modes. First, we considered a static credit network. In each step, the simulation executed a transaction and subsequently repaired the spanning tree if dynamic stabilization was applied. Afterwards, it returned the credit network to its original state. 
Second, we considered a dynamic network evolving over time. Transactions, node churn, and modifications of the extended credit changed
the structure of the network and the weights on the links. 
While the second mode was more realistic, it prevented a straightforward comparison of different approaches for individual transactions due to the differences in state at the time of the transaction.

We implemented the routing and stabilization algorithms of \cnsysname and \oursys as specified in Sections~\ref{sec:background} and \ref{sec:algo}, respectively. 
However, we disregard the cryptographic details
for our evaluation, as they do not affect our performance metrics. Instead, the \sender and \receiver both send only one message to each landmark forwarded by all nodes on the shortest path to the landmark. 
In our implementation of \cnsysname , each landmark then sends a message to all remaining landmarks, which is forwarded along the shortest paths, to account for the multi-party computation.
When combining embedding-based routing with multi-party computation, the \receiver sends messages to all landmarks.
In addition to enabling the evaluation of each individual modification, the alignment of the two designs also resulted in a fairer comparison of overheads, as the original \cnsysname sends all elements of a signature chain individually and thus results in a higher overhead as compared to sending them in one message.
As \cnsysname ' authors do not specify how the \sender decides on the amount of partial credit $c_i$ assigned to the \ith\ path, 
we decided to divide the total credit randomly between paths in agreement with the available minimum. In other words, if the sum of all minimal values was at least equal to the total transaction value $c$, we first divided $c$ randomly upon the paths.  We then randomly re-assigned all credit that exceeds the minimal value along a path to the remaining paths. We repeated the re-assignment step until the partial credit of each path was at most equal to the minimal credit on the path. 
During the simulation, we recorded all information necessary to derive the performance metrics described in Section~\ref{sec:model}.

For \oursys and \cnsysname , we consider the following parameters: i) the number of trees $\numT$, ii) the number of attempts $\numA$ that nodes try to perform a transaction before declaring it failed, iii) the maximal interval $\tl$ between two attempts for the same transaction, and iv) the interval $\epoch$
between two periodic re-computations of the trees for \cnsysname .
For comparison, we expressed the stabilization overhead for \oursys in stabilization messages per $\epoch$.
In addition to the above parameters, we provided two approaches for choosing landmarks: choosing the nodes of maximal degree or choosing random nodes.
Here, we define the maximal degree of a node as the number of connections with positive available credit in both directions. For the evolving credit network, we chose the nodes with the highest initial degree.

We implemented distributed versions of the
Ford-Fulkerson max-flow algorithm~\cite{ford1956maximal} and tree-only routing for comparison. Tree-only routing only uses links in the spanning tree but chooses the shortest path rather than always passing through the landmarks.  
For Ford-Fulkerson, we replaced the centralized computation with a distributed version that discovers residual flows using a breadth-first search. 
By adding tree-only routing, we evaluate all three tree-based routing schemes displayed in Figure \ref{fig:concepts-no-infer}, with SilentWhisper being an instance of landmark-centered routing and SpeedyMurmurs representing embedding-based routing.

\subsection{Dataset}
\label{sec:dataset}

We obtained datasets from crawling the PBT network Ripple~\cite{armknecht2015ripple}. In particular, we obtained crawls of the complete network from November 2016 and all link modifications and transactions since its creation in January 2013.
Based on these crawls, we derive datasets for both our simulation modes, the static and evolving network.
In the following, we first describe our crawling method, followed by post-processing of the crawled data. Last, we present properties of the resulting datasets.

\paragraph*{Dataset Processing} We restricted our evaluation to funded accounts: a Ripple account is {\em funded} when it owns a certain amount of XRP.\footnote{XRP is the symbol of the
Ripple currency.}  In April 2017, a user needed 20~XRP to fund an account. 
In this paper, we disregard transferring credit from one currency to another. Hence, we converted all values to US dollars and deleted all links and transactions in non-fiat currencies.
After cleaning the dataset according to these three rules, we derived the credit network $C'_{Nov16}$ for November 2016 and lists of both transactions and link value changes, sorted in temporal order. 
Based on the resulting transaction and link modifications lists, we then generated the credit network $C'_0$ at the time of the first transaction as a starting point of our second mode, the evolving network. 
As our data does not reveal when nodes join and leave the network, we included all crawled nodes and links in our initial credit network but set the weight of links $(u,v)$ that come into existence at a later point to $0$.
During the spanning tree construction, such links are ignored. 

We resolved three inconsistencies between our model and the datasets.
In rare cases, Ripple exhibits invalid credit arrangements; i.e., links $(u,v)$ such that their weight exceeds the upper limit of granted credit. Usually, such occurrences result from changes to the extended credit agreement. We deleted all such links from the dataset.  
Furthermore, we removed  self-transactions from the dataset, as they do not require routing algorithms according to our model.
Last, landmark routing requires paths between all nodes and the landmarks, so that we restricted our evaluation to the giant component.
These processing steps turned the initial snapshots $C'_{Nov16}$ and $C'_0$ into our processed datasets $C_{Nov16}$ and $C_0$. 
We obtained the final datasets by restricting the previous lists to entries involving only nodes in the final snapshots.

\paragraph*{Final Datasets} $C_0$ contained 93,502 nodes and a total of 331,096 links, whereas $C_{Nov16}$ contained 67,149 nodes and 199,574 links. The reason for the disparity is that $C_0$ contained all active links and their adjacent nodes for a period of more than 3 years, whereas $C_{Nov16}$ was a snapshot of the network on one particular date. 
Our final transaction lists had 970,472 and 692,737 entries for $C_0$ and $C_{Nov16}$, respectively.  We recorded a total of 652,216 link modifications for the evolving network $C_0$. 
The datasets and the code are publicly available.\footnote{\url{https://crysp.uwaterloo.ca/software/speedymurmurs/}}

\begin{table*}[t]
\centering
\begin{small}
\caption{Performance of different transaction schemes in the static scenario, varying the routing algorithm (LM-Landmark, GE-greedy embedding, TO-Tree-only), the stabilization method (PER-periodic, OND-on-demand), the assignment of credit on paths (MUL-multi-party computation, RAND-random), and the landmark selection (HD-highest degree, RL-random landmark) for five metrics: success ratio: fraction of successful transactions (higher is better), delay: longest chain of messages (lower is better), transaction: messages sent per transaction (lower is better), path length: length of discovered paths between sender and receiver (lower is better), stabilization: messages for stabilizing the trees sent per epoch (lower is better).
SilentWhispers corresponds to the setting LM-MUL-PER whereas SpeedyMurmurs is GE-RAND-OND.}
\begin{tabular}{@{}l|d{3}@{\hskip 0.05cm}c@{\hskip 0.2cm}d{3}|d{3}@{\hskip 0.05cm}c@{\hskip 0.2cm}d{3}|d{2}@{\hskip 0.05cm}c@{\hskip 0.2cm}d{2}|d{3}@{\hskip 0.05cm}c@{\hskip 0.2cm}d{3}|d{0}@{\hskip 0.2cm}c@{\hskip 0.2cm}d{0}}
Setting & \multicolumn{3}{c|}{Success Ratio} & \multicolumn{3}{c|}{Delay} & \multicolumn{3}{c|}{Transaction} &   \multicolumn{3}{c|}{Path Length} & \multicolumn{3}{c}{Stabilization}\\ 
 & \multicolumn{3}{c|}{} & \multicolumn{3}{c|}{(Hops)} & \multicolumn{3}{c|}{(Messages)} &   \multicolumn{3}{c|}{(Hops)} & \multicolumn{3}{c}{(Messages)} \\ \hline
\rowcolor{Gray}
\textbf{SilentWhispers-HD} & 0.651 & $\pm$ &  0.005 & 15.01 & $\pm$ &  0.08 & 82.0 & $\pm$ &  0.2 & 5.30 & $\pm$ &  0.01 & 598722 & $\pm$ &  0 \\
LM-MUL-OND-HD & 0.62 & $\pm$ & 0.03 & 14.7 & $\pm$ & 0.5 & 81 & $\pm$ & 2 & 5.3 & $\pm$ & 0.1 & 8000000 & $\pm$ & 2000000\\
\rowcolor{Gray}
LM-RAND-PER-HD & 0.09 & $\pm$ & 0.01 & 8.3 & $\pm$ & 0.1 & 35.1 & $\pm$ & 0.5 & 3.23 & $\pm$ & 0.05 & 598722 & $\pm$ & 0\\
LM-RAND-OND-HD & 0.09 & $\pm$ & 0.09 & 9 & $\pm$ & 1 & 37 & $\pm$ & 4 & 3.4 & $\pm$ & 0.4 & 2000 & $\pm$ & 2000\\
\rowcolor{Gray}
GE-MUL-PER-HD & 0.908 & $\pm$ & 0.001 & 11.52 & $\pm$ & 0.03 & 49.0 & $\pm$ & 0.1 & 1.951 & $\pm$ & 0.003 & 598722 & $\pm$ & 0\\
GE-MUL-OND-HD & 0.905 & $\pm$ & 0.004 & 11.5 & $\pm$ & 0.2 & 49.0 & $\pm$ & 0.5 & 1.954 & $\pm$ & 0.007 & 4000 & $\pm$ & 4000\\
\rowcolor{Gray}
GE-RAND-PER-HD & 0.913 & $\pm$ & 0.001 & 6.016 & $\pm$ & 0.009 & 18.30 & $\pm$ & 0.04 & 1.867 & $\pm$ & 0.003 & 598722 & $\pm$ & 0\\
\textbf{SpeedyMurmurs-HD} & 0.906 & $\pm$ & 0.006 & 6.02 & $\pm$ & 0.04 & 18.3 & $\pm$ & 0.1 & 1.87 & $\pm$ & 0.01 & 300 & $\pm$ & 300\\
\hline 
\rowcolor{Gray}
TO-SW-HD & 0.863 & $\pm$ & 0.003 & 15.9 & $\pm$ & 0.1 & 81.9 & $\pm$ & 0.3 & 3.17 & $\pm$ & 0.01 &598722 & $\pm$ & 0\\
TO-SM-HD & 0.54 & $\pm$ & 0.04 & 6.7 & $\pm$ & 0.3 & 23.5 & $\pm$ & 0.7 & 2.01 & $\pm$ & 0.07 & 5000 & $\pm$ & 5000\\
\rowcolor{Gray}
Ford-Fulkerson & 1.00 & $\pm$ & 0.00 & 49500 & $\pm$ & 900 & 49500 & $\pm$ & 900 & 3.2 & $\pm$ & 0.1 & 0 & $\pm$ & 0\\
\hline
SilentWhispers-RL & 0.1 & $\pm$ & 0.2 & 15 & $\pm$ & 2 & 130 & $\pm$ & 10 & 7.6 & $\pm$ & 0.6 & 598722 & $\pm$ & 0\\
\rowcolor{Gray}
SpeedyMurmurs-RL & 0.912 & $\pm$ & 0.007 & 5.99 & $\pm$ & 0.06 & 18.2 & $\pm$ & 0.2 & 1.863 & $\pm$ & 0.009 & 1000  & $\pm$ & 1000\\
\end{tabular}
\label{tab:algos}
\end{small}

\end{table*}

\subsection{Simulation Setup}

Our first simulation setup realized the static simulation mode on the basis of the snapshot $C_{Nov16}$. 
We repeated simulations 20 times, using a different set of 50,000 transaction for each run.
We chose these transactions pseudorandomly, seeded by the run number, from all transactions that were successful using Ford-Fulkerson, a total of 331,642 transactions. 
We then evaluated all 8 possible combinations of routing algorithms (landmark routing or embedding-based), credit assignments to paths (multi-party computation or random assignment), and stabilization algorithms (periodic or on-demand) for the parameters $\numT=3$ and $\numA=2$.  
We chose $\epoch=1000$, meaning we recomputed spanning trees each 1000 transactions. We choose the re-queuing interval as $\tl=2\cdot\epoch$. For the landmark selection, we considered both options: random and highest degree. Note that random choices were deterministic in the run number, ensuring comparability of all approaches under the same circumstances.
For comparison with related approaches, we evaluated two versions of tree-only routing, using \cnsysname' multi-party computation and periodic stabilization for the first version and \oursys ' random credit assignment and on-demand stabilization for the second. 
We then evaluated the impact of the different parameters for \cnsysname and \oursys . We vary the number of landmarks $\numT$ between 1 and 7 and the number of attempts $\numA$ between 1 and 10.

Our second simulation setup realized the evolution of the network under different algorithms: Ford-Fulkerson, \cnsysname , and \oursys .
Starting from the initial network $C_0$, the simulation initiated the transactions and changes link values according to the dataset.
For \cnsysname and \oursys , we set $\numT=3$, $\numA=2$, $\epoch=1000\delta_{Av}$, and $\tl=2\delta_{Av}$
with $\delta_{Av}$ denoting the average time between two transactions. In this manner, an epoch roughly corresponds to a day. 
We chose landmarks of the highest degree for \oursys and \cnsysname.
As Ford-Fulkerson is a deterministic algorithm, we only executed it once but averaged our results for \oursys and \cnsysname over 20 runs.

\subsection{Results}

We start by comparing a wide range of algorithms for the static simulation setup.
Table~\ref{tab:algos} displays the results for different combinations of the three proposed modifications to \cnsysname  as well as our implementations of tree-only routing and Ford-Fulkerson. 
Note that Ford-Fulkerson is a deterministic algorithm but its delays and overheads vary as the set of transactions varies between runs.

\paragraph*{Impact of Design Decisions} 
As expected, greedy embeddings led to shorter paths due to finding shortcuts between different branches of the tree. Hence, all settings using greedy embeddings, i.e., rows starting in ``GE-'' and SpeedyMurmurs, exhibited lower delays and transaction overheads than the corresponding landmark-based algorithms. Indeed, greedy embeddings reduced the path length and the transaction overhead by nearly a factor of 2.
Greedy embeddings also increased the success ratio due to the shorter paths and the lower probability of encountering a link with low available credit.

The impact of the random assignment of credit, used by SpeedyMurmurs and all algorithms with ``RAND'' in their name, was less clear-cut: While removing the need to involve landmarks into the routing process reduced the delay and the transaction overhead for all parameter settings, the impact on the success ratio differed between embedding-based and landmark routing. 
When combined with landmark routing, random credit assignments resulted in a definite drop in success from more than 60\% to only 8\%. The reason for the low success ratio was the high probability of encountering at least one link with insufficient credit to satisfy the random assignment.
In contrast, greedy embeddings exhibited much shorter paths and the flexibility to potentially choose between several neighbors. These two properties negated the disadvantageous impact of the random credit assignment, so that greedy embedding in combination with random assignment resulted in the same success ratio of 91\% as in combination with multi-party computation. 

On-demand stabilization reduced the stabilization overhead (abbreviated by Stabilization in Table~\ref{tab:algos})
drastically: While rebuilding the spanning trees periodically, as applied by SilentWhispers and all algorithms with ``PER'' in their name, resulted in more than half a million messages per epoch, on-demand stabilization only required a few thousands of messages,
as shown in the last column of Table~\ref{tab:algos}. 
On-demand stabilization induced high variance because the value of links close to the root of a spanning tree rarely drops to 0 but incurred enormous overhead in these rare occurrences. 
The simulation showed a clear advantage of on-demand stabilization. 
We admit that the considerable advantage of on-demand stabilization is partially due to lack of link value changes and actual dynamics in the static simulation. In the second part of this section, we therefore evaluate the stabilization overhead in a dynamic environment.

\begin{figure*}[t]
\centering
\subfloat[Success Ratio : Trees (Higher is Better)]{\label{fig:treesSR}\includegraphics[width=0.33\linewidth]{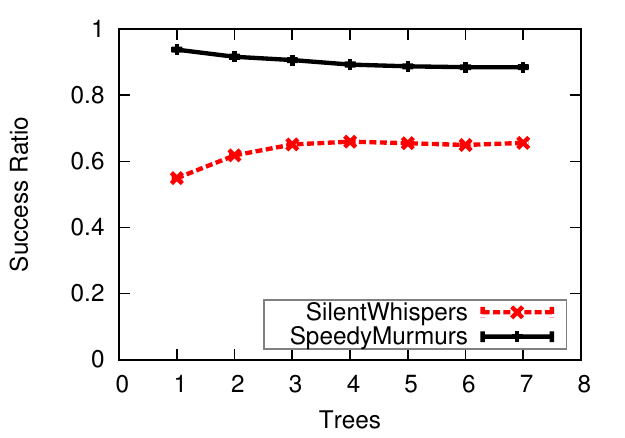}}
\subfloat[Delay : Trees (Lower is Better)]{\label{fig:treesDelay}\includegraphics[width=0.33\linewidth]{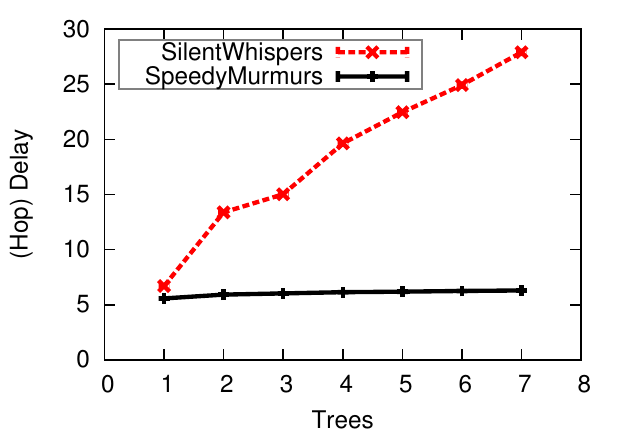}}
\subfloat[Success Ratio : Attempts (Higher is Better)]{\label{fig:attSR}\includegraphics[width=0.33\linewidth]{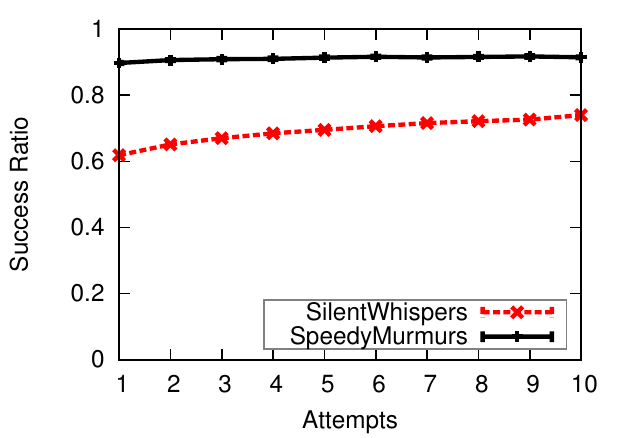}}
\caption{\cnsysname vs.\ \oursys: Impact of number of parallel trees and attempts at performing a transaction}
\label{fig:parameter}
\vspace{-2em}
\end{figure*}

\begin{figure*}[t]
\centering
\subfloat[Events/Epoch]{\label{fig:events}\includegraphics[width=0.33\linewidth]{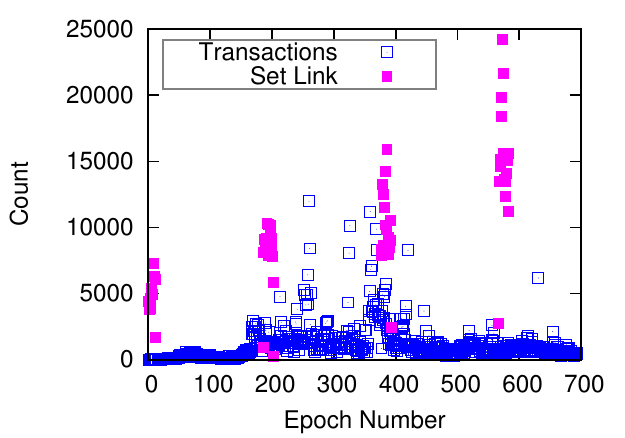}}
\subfloat[Stabilization (Lower is Better)]{\label{fig:stab}\includegraphics[width=0.33\linewidth]{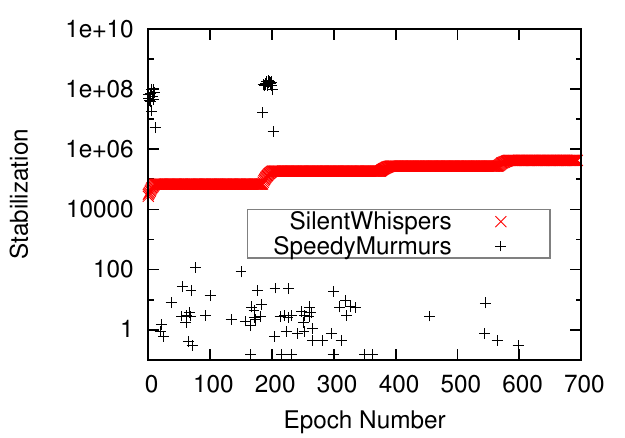}}
\subfloat[Success (Higher is Better)]{\label{fig:succ}\includegraphics[width=0.33\linewidth]{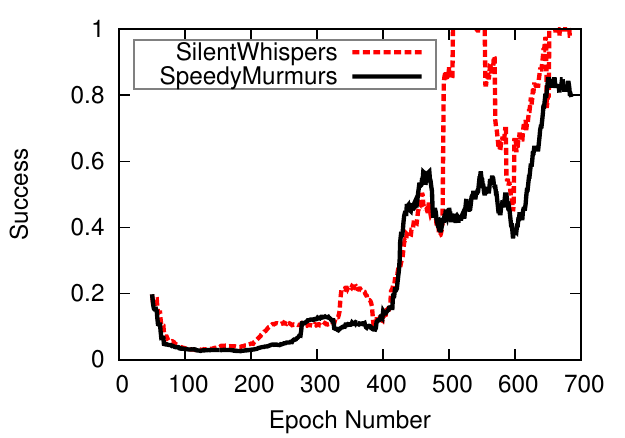}}
\caption{Comparing \oursys and \cnsysname in a dynamic setting based upon Ripple transaction and link changes from 2013 to 2016 on per-epoch scale; success is computed as the ratio of the actual success ratio and the success ratio of the Ford-Fulkerson algorithm as a baseline; for c),  we present moving averages over 50 epochs to increase readability}
\label{fig:dynamic}
\vspace{-1em}
\end{figure*}

\paragraph*{Comparison to Other Algorithms} We compared \cnsysname and \oursys with tree-only routing and Ford-Fulkerson based on the results in Table~\ref{tab:algos}.
As expected, Ford-Fulkerson exhibited prohibitive delays and transaction overheads. The fact that Ford-Fulkerson also results in a longer average path length seems counterintuitive at first. However, the result was a side effect of Ford-Fulkerson discovering long paths to maximize the available credit that the other approaches failed to discover. 
As illustrated in Fig.~\ref{fig:concepts-no-infer} and detailed in Sec.~\ref{sec:background}, tree-only routing finds the shortest route in the spanning tree,
possibly without passing a landmark, but does not include links that are not contained in the tree.
Thus, tree-only routing is a compromise between \cnsysname' routing algorithm and embedding-based routing. 
As a consequence, the performance results when using only tree links are in between the performance of \cnsysname and \oursys .

\paragraph*{Impact of $\numT$ and $\numA$} Next, we evaluate the impact of different configuration parameters on the performance. 
As indicated in the last two rows of Table~\ref{tab:algos}, choosing random landmarks did not considerably affect the performance of \oursys but reduced the performance of \cnsysname due to the existence of longer paths to a landmark with few connections. 
In contrast, increasing the number of trees $\numT$ affected the success ratio of \oursys negatively and \cnsysname positively, as Fig.~\ref{fig:treesSR} indicates. 
The reason for the observed decrease in success was the increased likelihood that at least one path did not have sufficient credit. 
An increased $\numT$ further increased the delays, as shown in Fig.~\ref{fig:treesDelay}. The impact was more pronounced for \cnsysname because landmarks had to wait until all messages for the multi-party computation arrived.   
The number of attempts $\numA$ had a slight positive effect on the success ratio, as shown in Fig.~\ref{fig:attSR}. Yet, as the transaction overhead is linear in the number of attempts, the slight increase may not warrant multiple attempts.  

For all algorithms but Ford-Fulkerson, the success ratio was considerably below 100\%. It stands to reason that a lot of users might not be willing to accept a failure rate of 10\% or more. Note that a failure to route does not reduce the funds of any user, so there is no loss in funds associated with a routing failure.  
Furthermore, in a non-static environment, users can retry the transaction at a later point in time after the network has sufficiently changed for it to work.  
If neither failure nor waiting is an option, we could apply Ford-Fulkerson on failure. By reducing the transactions that require Ford-Fulkerson to 10\%, we still considerably improve the efficiency in comparison to a network relying exclusively on Ford-Fulkerson at the price of a slight increase in delay due to the preceding use of \oursys. 
In addition, we hope that with increasing popularity, both the connectivity of the PBT networks and the amount of available funds increase beyond the current state of the Ripple network, which is bound to entail a higher probability of success.

\paragraph*{Impact of Dynamics} We evaluated the impact of dynamics on the performance of \oursys and \cnsysname . As stated above, the impact of dynamics is particularly of interest to decide if on-demand stabilization is indeed more efficient than periodic stabilization. 
To better comprehend the reasons underlying our results, Fig.~\ref{fig:events} displays the number of transactions and link changes per epoch for the Ripple dataset.
While the number of transactions did not vary greatly over the period of three years, link creations and modifications were frequent in some short intervals but rare during the remaining observation period. The frequency of link changes directly relates to the stabilization overhead of \oursys , as indicated by Fig.~\ref{fig:stab}. Whereas the stabilization overhead was usually below 100 messages per epoch, the overhead increased to about $10^9$ messages during periods of frequent change. Note that only the first two of the four batches of link changes resulted in a drastically increased need for stabilization. After the first two batches, spanning trees had formed and new link additions mostly created shortcuts that did not require changes to the trees. 
In contrast, the stabilization overhead of \cnsysname only depended on the number of edges in the network and hence increased as the graph grows over time. During intervals of frequent change, the stabilization overhead of \cnsysname was considerably lower than \oursys ' stabilization overhead. However, during `normal' operation, \cnsysname' stabilization overhead exceeded the overhead of \oursys by more than 2 orders of magnitude.  
We evaluated the success in relation to Ford-Fulkerson and hence  divided the actual success ratio of each epoch by the success ratio of Ford-Fulkerson for the corresponding epoch. 
As can be seen from Fig.~\ref{fig:succ}, the success could exceed 1 if an alternative routing algorithm exhibited a higher success ratio. Note that higher success ratios were indeed possible due to the fact that different routing algorithms resulted in different payments and hence different network states. Different network states implied a different set of possible transactions, so that a transaction could fail for Ford-Fulkerson but succeed for \cnsysname or \oursys. 
In comparison, \oursys and \cnsysname achieved similar success ratios for most of the time; however, at the end of the simulation interval, \cnsysname outperformed \oursys . The sudden increase in success correlates with the addition or change of many links, as can be seen from Fig.~\ref{fig:events}. The additional links increase the density of the graph, leading to shorter paths, and hence a higher success probability. The fact that \cnsysname achieves a higher success ratio than \oursys could be due to the tree structure: \cnsysname maintains breadth-first search trees whereas \oursys initially constructs breadth-first search trees but does not change the parent of a node if a new neighbor offers a shorter path to the root. The longer paths to the root could have negative effects on the probability of success. 
As the actual success ratio of all considered algorithms is low during later epochs, e.g., frequently below $5$\%, the result might be an artifact of our dataset and post-processing method. 

We hence answered our five initial research questions: 
\begin{itemize}
\item \oursys achieved a higher performance than \cnsysname with regard to all considered metrics for the static  scenario. 
\item On-demand stabilization and embedding-based routing had a positive effect on all 5 performance metrics. In contrast, the use of random credit assignment might decrease the success ratio slightly. However, when used in combination with embedding-based routing, the effect was mostly negated.
\item As expected, Ford-Fulkerson usually achieved a higher success than both \oursys and \cnsysname . However, the algorithm resulted in an enormous transaction overhead, exceeding the overhead of the other algorithms by 2 to 3 orders of magnitude. 
\item An increased number of trees or attempts to perform a transaction did not considerably increase the success ratio  of \oursys but incurred increased overheads.
\item The evolution of the PBT network affects the performance of SpeedyMurmurs considerably. Stabilization overhead and success ratio vary considerably depending on the frequency of transactions and link changes. 
\end{itemize}
 The dynamic evaluation suggests working on the design of an alternative spanning tree maintenance algorithm.
 In particular, the results raise the question of suitable criteria for dynamically switching between on-demand and periodic stabilization. Indeed, as \cnsysname is more efficient during periods of frequent change but results in higher overhead otherwise, such a switching mechanism could further reduce the communication overhead and hence increase scalability.


\section{Related Work}
\label{sec:related}

Maximizing the set of possible transactions in a credit network is NP-hard~\cite{Ghosh07}. 
Instead, many existing systems have opted for considering one transaction at a time and 
applying the max-flow approach~\cite{ford1956maximal} as a routing algorithm.   
Nevertheless, existing algorithms~\cite{max-flow-v2} run in $O(V^3)$ or $O(V^2 \log(E))$ time and 
hence do not scale to a growing number of \users and transactions~\cite{viswanath2012canal, post2011bazaar}.

The pioneering credit networks Ripple and Stellar maintain their entire PBT networks on public blockchain ledgers. 
Although this information can be leveraged to 
perform routing efficiently, it also trivially leaks sensitive information such as 
credit links/relationships and financial activity in the form of transactions.  Instead, current proposals rely  
on a decentralized PBT network requiring no public log.

Prihodko et al.\ recently proposed Flare~\cite{prihodko2016flare}, a routing algorithm for the 
 Lightning Network, 
a network of Bitcoin payment channels among Bitcoin \users that enables off-chain transactions~\cite{poon2015bitcoin}.
In Flare, all nodes keep track of their $k$-neighborhood; i.e., nodes at a hop distance of at most $k$ and all links between them. In addition, each node maintains paths to a set of nearby \emph{beacon} nodes.

This routing algorithm reveals the weight of all links in the $k$-neighborhood, usually for $k\geq 3$. 
This results in
a privacy concern as the weight of a link between two users is exposed to users other than those two.
Furthermore, nodes spread all updates to the $k$-neighborhood, meaning each credit change results in possibly hundreds of messages, which is highly inefficient for frequent transactions and hence changes in available credit.

Canal~\cite{viswanath2012canal} presents the first efficient implementation of tree-only routing applied to 
looking for paths in credit networks. A trusted central party computes the shortest paths in the spanning trees between sender and receiver. If these paths provide enough credit to settle a transaction, the routing terminates successfully. Otherwise, it fails. 
In the face of network dynamics, the central server re-computes spanning trees constantly. 
Due to maintaining a central server, Canal~\cite{viswanath2012canal} has severe privacy and security drawbacks.

PrivPay~\cite{moreno15privpay} increases the privacy of Canal by using trusted hardware at the central server. However, PrivPay relies on a 
similar landmark technique as Canal and is also a centralized solution, therefore the scalability is still low and the issue of a single point of failure remains unsolved. 
Additionally, the PrivPay paper introduces for the first time the notions of 
value privacy and sender/receiver privacy for payments in a credit network.
In this work, we define the privacy notions for routing in a \paysys as a building 
block not only for credit networks but also for any \paysys.

SilentWhispers~\cite{malavolta17silent} uses landmark routing in a fully distributed credit network. 
Both \sender and \receiver send messages in the direction of the landmarks, 
which constitute rendezvous nodes. In other words, 
paths in SilentWhispers are concatenations of the \sender's path to a landmark and the path 
from said landmark to the \receiver. 
All paths pass a landmark, even if \sender and \receiver happen to be in the same branch, potentially leading to performance issues. However, as we discuss throughout this paper, SpeedyMurmurs, the routing 
algorithm proposed in this work, outperforms the routing approach proposed in SilentWhispers while 
achieving the privacy notions of interest.


Malavolta et al.~\cite{malavolta17PCN}
recently proposed Rayo and Fulgor, two payment-channel 
networks (i.e., \paysyss) 
that provide a necessary tradeoff between privacy and concurrency. Their  
study of concurrency could be leveraged to extend how 
concurrency is handled in \oursys. Nevertheless, they
do not tackle the path selection problem. Thus, 
\oursys is an excellent candidate to complement Rayo and Fulgor.

\subsubsection*{Summary} Existing routing approaches often disregard privacy. Most of them require centralization or shared public information; SilentWhispers is the only existing distributed PBT network focusing on privacy. However, it relies on a distributed landmark routing technique that is potentially inefficient. Our in-depth performance and privacy evaluation 
shows that SpeedyMurmurs provides higher overall 
performance 
when compared to state-of-the-art routing approaches, while 
achieving the privacy notions of interest.

\section{Conclusion and Future Work}
\label{sec:conc}

In this work, we design \oursys, an efficient routing algorithm for completely decentralized PBT networks.
Our extensive simulation study and analysis indicate that \oursys is highly efficient and achieves a high probability of success
while still providing value privacy as well as \sender/\receiver privacy against a strong network adversary.
As these privacy notions are essential for PBT applications,
\oursys is an ideal routing algorithm for decentralized credit networks and payment channel networks, as well as for emerging inter-blockchain algorithms.

As our results indicate that on-demand and periodic stabilization are suitable for different phases of a PBT network's evolution, future work can extend upon our results by investigating the option of dynamically switching between on-demand and periodic stabilization. 

\vspace{-1em}

\section*{Acknowledgements}
This work benefited from the use of the CrySP RIPPLE Facility at 
University of Waterloo and is partially supported by an Intel/CERIAS RA and NSERC grant
RGPIN-2017-03858. 

\vspace{-1em}

\bibliographystyle{plain}
\balance
\bibliography{bibliography}

\end{document}